\documentclass[twocolumn,fontsize=10]{aastex61}
\usepackage{amsmath,color,url}
\usepackage[version=4]{mhchem}

\begin{document}

%\title{Convective dynamics and disequilibrium chemistry in the atmospheres of (self-luminous?) giant planets and brown dwarfs}
\title{Convective dynamics and disequilibrium chemistry in the atmospheres \\ of giant planets and brown dwarfs}

%% Use \author, \affil, plus the \and command to format author and affiliation 
%% information.  If done correctly the peer review system will be able to
%% automatically put the author and affiliation information from the manuscript
%% and save the corresponding author the trouble of entering it by hand.
%%
%% The \affil should be used to document primary affiliations and the
%% \altaffil should be used for secondary affiliations, titles, or email.

%% Authors with the same affiliation can be grouped in a single
\author{Baylee Bordwell\altaffiliation{1} and Benjamin P. Brown\altaffiliation{1}}
\affil{Department of Astrophysical and Planetary Sciences,\\
  Laboratory for Atmospheric and Space Physics, University of Colorado at Boulder\\
  3665 Discovery Drive \\
  Boulder, CO 80303-7814, USA}
\email{baylee.bordwell@colorado.edu}
%% Use the \and command so offset the last author.
%\and
\author{Jeffrey S. Oishi\altaffiliation{2}}
\affil{Department of Physics and Astronomy\\
  Bates College\\
  Carnegie Science Hall\\
  2 Andrews Road\\
  Lewiston, ME, 04240, USA}

%\altaffiltext{1}{AAS Journals Data Scientist}
%\altaffiltext{2}{greg.schwarz@aas.org}
%\altaffiltext{3}{AAS Journals Associate Editor-in-Chief}
%\altaffiltext{4}{AAS Director of Publishing}
%\altaffiltext{5}{IOP Senior Publisher for the AAS Journals}

\begin{abstract}
  Disequilibrium chemical processes have a large effect upon the spectra of substellar objects.
  To study these effects, dynamical disequilibrium has been parameterized using the quench and
  eddy diffusion approximations, but little work has been done to explore how these approximations
  perform under realistic planetary conditions in different dynamical regimes. As a first step
  in addressing this problem, we study the localized, small scale convective dynamics of planetary
  atmospheres by direct numerical simulation of fully compressible hydrodynamics with reactive
  tracers using the Dedalus code.
  Using polytropically-stratified, plane parallel atmospheres in 2- and 3-D, we explore the quenching
  behavior of different abstract chemical species as a function of the dynamical conditions
  of the atmosphere as parameterized by the Rayleigh number. We find that in both 2- and 3-D, chemical
  species quench deeper than would be predicted based on simple mixing length arguments.
  Instead, it is necessary to employ length scales based on the chemical equilibrium profile of the
  reacting species in order to predict quench points and perform chemical kinetics modeling in 1-D.
  Based on the results of our simulations, we provide a new length scale,
  derived from the chemical scale height, which can be used to perform these calculations. This
  length scale is simple to calculate from known chemical data and makes reasonable predictions
  for our dynamical simulations.
\end{abstract}

%% See the online documentat ion for the full list of available subject



\section{Introduction} \label{sec:intro}

As the number of known substellar objects has grown into the thousands,
a unique testing ground has developed for our understanding of planetary
atmospheres. Substellar objects with masses similar to or in
excess of that of Jupiter, subject to little or no irradiation,
comprise an especially interesting sample for study,
yielding strong observational signals while simultaneously affording relatively
simple atmospheric chemistry and dynamics for modeling \citep{Moses2014,Showman2010}.
%\footnote{
%  Would you say that the dynamics of a substellar
%        object are easier to model than those of a Tellurian atmosphere? why or why not?
%        I'm thinking along the lines of the lack of a lot of geologic feedbacks...}
Jupiter, directly imaged giant exoplanets, and brown dwarfs all
share a similar general atmospheric structure, with a lower convective zone
driven by cooling of the interior and a cool overlying radiative zone \citep{Burrows1997,
  Showman2010}.
%This structure facilitates both the generation of especially deep
%molecular features \citep{Madhusudhan2014} and the occurrence of transport-induced
%disequilibrium chemistry under certain conditions \citep{Prinn1977,Saumon2003,Saumon2007,
%  Zahnle2014,Moses2016}.
This structure can facilitate the generation of especially strong molecular features
\citep{Madhusudhan2014}, which has led to the (unambiguous) detection of 
 CH$_4$, H$_2$O, and CO in the atmospheres of directly imaged giant exoplanets
\citep{Patience2010,Barman2011,Barman2015,Oppenheimer2013,Konopacky2013,Janson2013,
  Snellen2014,Chilcote2015,Macintosh2015,Samland2017}
and CH$_4$, NH$_3$, H$_2$O, CO and CO$_2$ in the atmospheres of brown
dwarfs \cite[e.g.,][]{Geballe1996,Geballe2009,Oppenheimer1995,Oppenheimer1998,Schultz1998,
  Noll1997,Noll2000,Leggett2000,Saumon2000,Saumon2007,Yamamura2010}. In particular, the
abundances of these molecules are sensitive to the strength of vertical mixing
%\footnote{This gets
%  tricky with ammonia, so we may get called out on this statement}
in these atmospheres,
making it essential to accurately incorporate the effect of dynamics into the interpretation
of these observations \citep{Saumon2003,Zahnle2014,Moses2016}.

To explain  observations of CO in the atmosphere of Jupiter, \citet{Prinn1977} (hereafter PB77)
used mixing length theory to describe atmospheric transport,
and found that in the limit of small chemical scale
height (the scale height of the chemical reaction rate), the predicted abundance will quench to the equilibrium value
at the pressure where chemical and dynamical timescales become equal.
This approximation provides a framework for tying the strength of atmospheric mixing to
observed abundances in the form of the eddy diffusion coefficient,
\begin{equation}\label{eddy_diff}
  K_{zz} = w(z)L(z),
\end{equation} 
where $w$ is a vertical velocity and L(z) is the characteristic length scale of the atmosphere,
which in PB77 is taken to be a density scale height. Further work by \citet{Smith1998} demonstrated
that the characteristic length scale for quenching is a function of properties of the equilibrium profile and
chemical and dynamical timescales. Modern 1-D atmospheric models utilizing advanced
chemical kinetics (e.g., \citet{Moses2011}) build upon this work by directly employing $K_\text{zz}$ in the
evolution equation for each chemical species,
\begin{equation}\label{chemev}
  \frac{\partial n_i}{\partial t} = \frac{\partial}{\partial z}\left[
  \left(K_{zz}+D_{i}\right)\frac{\partial n_i}{\partial z}\right] + \mathcal{P}_i - \mathcal{L}_i,
\end{equation}
where $n_i$ is the concentration, or number density, of species $i$,
$D$ is the molecular diffusion coefficient,
and $\mathcal{P}_i$ and $\mathcal{L}_i$ are production and loss rates \citep{Moses2011}. To estimate
profiles of $K_\text{zz}$ for these models, a combination of analytical approximations and
simulated velocity profiles from 3-D general circulation models (GCM) are typically used.
While these 1-D models have demonstrated the presence of quenching behavior for all of the species discussed
above in the atmospheres of Jupiter, directly imaged giant exoplanets, and brown dwarfs
\cite[e.g.,][]{Moses2005,Visscher2011,Zahnle2014,Moses2016}, they are fundamentally
limited by their dependence upon accurate prescriptions for the dynamics.

Various efforts have been made to explore and describe the transport properties of these atmospheres
in 2- and 3-D simulations that employ more realistic dynamics using passive and reactive
tracers. One approach has been to include the evolution of a reactive \citep{Cooper2006}
or settling passive \citep{Parmentier2013} tracer to a 3-D GCM solving the primitive
equations in the radiative zone. While \citet{Cooper2006} found that quenching occurred
as predicted based on \citet{Smith1998}, \citet{Parmentier2013} saw that measured eddy
diffusion coefficients differed from those estimated conventionally using Equation
(\ref{eddy_diff}) by orders of magnitude. Another approach, taken by \citet{Freytag2010}
has been to consider the evolution of a reactive tracer in a 2-D local box model solving
the fully compressible equations of hydrodynamics in a coupled radiative-convective atmosphere.
\citet{Freytag2010} observed similar discrepancies between measured and estimated eddy
diffusion coefficients. Given these discrepancies, it worth a closer look into the
applicability of the eddy diffusion approximation in chemical mixing, and the current
models for chemistry in dynamic disequilibrium.

In the present work, we take a different approach from previous dynamics studies. Using 2-D and
3-D local convective box models, we hold the majority of atmospheric parameters constant,
and explore the transport properties of a reactive tracer as a function of increasingly realistic
dynamical forcing regimes. These simple convection experiments are the regime for which the mixing
behavior should be well-described by
mixing length theory and eddy diffusion. We compare our results to the predictions of PB77 and
\citet{Smith1998}, as well as a 1-D model employing a vertical eddy diffusion coefficient profile
determined using the results of our simulation. We find an agreement with \citet{Smith1998} that
our measured quench points are significantly deeper than would be predicted by PB77, in both 2- and
3-D. We find similarly good agreement with a length scale, $H_\text{chem,eq}$, in inferring the quench point from data and predicting it from 1-D models. This length scale is simple to calculate from known chemical data. We conclude with a discussion of our results and their implications for the modeling of chemical disequilibrium.

\section{2- and 3-D Models} \label{sec:methods}
\begin{figure*}
  \centering
  \includegraphics[width=7in]{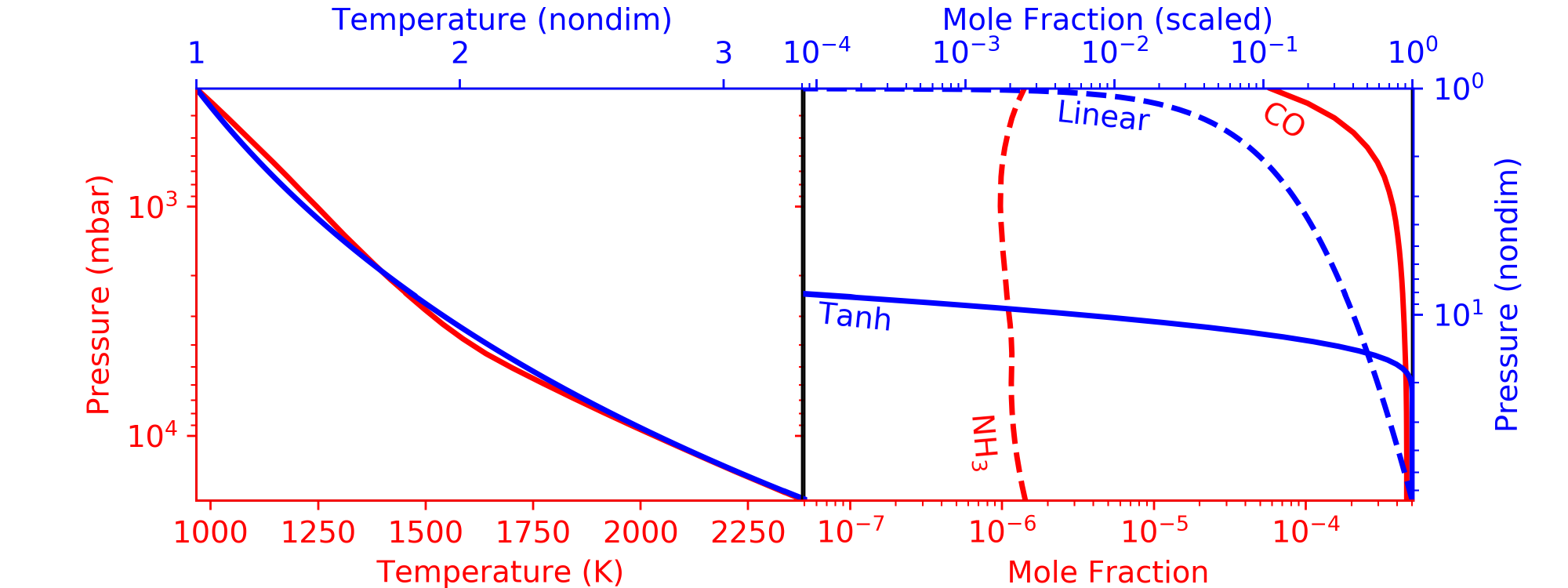}
  \caption{Simulation temperature-pressure profile and equilibrium profiles (blue) compared with those across
    the convecting region of a 1000K directly imaged giant exoplanet from \citet{Moses2016} (red). The CO
    equilibrium profile shown here is an example of a rapidly varying, (e.g., similar to a tanh), profile, while a
    linear profile might be more appropriate for a molecule like NH$_3$.\label{setup}}
\end{figure*}
\subsection{Equations and assumptions}
We solve the fully compressible equations of hydrodynamics,
  \begin{subequations}
\footnotesize
  \begin{align}
    \frac{\partial}{\partial t}(\rho) + \pmb{\nabla\cdot}(\rho\mathbf{u}) &= 0,\\
    \frac{\partial}{\partial t}(\rho\mathbf{u}) + \pmb{\nabla\cdot}(\rho\mathbf{u}\mathbf{u}) &=
    -\rho \mathbf{g} +\pmb{\nabla\cdot}\sigma,\\
    \frac{\partial }{\partial t}(\rho e) + \pmb{\nabla\cdot}(\rho e\mathbf{u})  &=
    \pmb{\nabla\cdot}(\chi\rho\pmb{\nabla} T)+(\sigma\pmb{\cdot\nabla})\pmb{\cdot}\mathbf{u},
  \end{align}
  \end{subequations}
  where $\rho$, $\mathbf{u}$, $\sigma$, $e$, and $T$ are the fluid mass density, velocity,
  stress tensor, specific internal energy, and temperature respectively, $g$ is the gravitational
  acceleration, and $\chi$ is the thermal diffusivity. We make the assumption of a
  Newtonian stress tensor, with constant dynamic viscosity $\mu = \nu\rho$, and a constant
  Prandtl number,
  \begin{equation}
    \text{Pr} = \frac{\nu}{\chi} = 1.
  \end{equation} The viscosity is defined in terms
  of the Rayleigh and Prandtl numbers as,
    \begin{equation}
      \nu = \sqrt{\text{Pr}\frac{gL_z^3\Delta s/c_p}{\text{Ra}}},
    \end{equation}
    where $L_z$ is the depth of the convective zone, $\Delta s$ is the entropy jump across the domain,
    and $c_p$ is the specific heat capacity at constant pressure. Rayleigh numbers specified in this work
    are defined at the top of the atmosphere, but increase by a factor of $e^{2n_H}$ (where $n_H$ is the
    number of scale heights across the domain) from top to bottom.
    The Rayleigh number is left as a free parameter that is explored in this study, as realistic Rayleigh
    numbers for planetary atmospheres ($\approx 10^{20}$) are not within reach of current simulations.
    %\footnote{
    %  Callout other work on Ra? Make a table with our Ra range (defined at middle of domain, divided by
    %  critical), theirs (all divided by critical), and realistic values for DIEGP and Jupiter}
    We hold gravitational acceleration to be constant, but in our formulation of these equations,
    we do not make the hydrostatic approximation, which can compromise the accuracy of interpretation
    of phenomena which occur on smaller horizontal scales \citep{Mendonca2016}.

  We assume an ideal gas equation of state
  with a ratio of specific heats $\gamma=7/5$, to emulate a primarily diatomic molecular gas. 
  Our initial conditions and background are of a polytropic atmosphere. For a convective region,
  this has a polytropic index of,
  \begin{equation}
    m = \frac{1}{\gamma-1} - \epsilon,
  \end{equation}
  where $\epsilon$ is the superadiabatic excess, which we take to be $10^{-4}$ based on work
  by \citet{Anders2017} demonstrating that $\epsilon\propto \text{Ma}^2$ and based on MLT calculations
  finding that typical Ma for these objects should be $\sim10^{-2}$.

  Polytropic atmospheres have
  background temperature profiles ($T_0$) that are linear with altitude,
  which leads to the profile shown in the left panel of Figure \ref{setup}.
  For an ideal gas, as we work with here, with $P=\rho T$, the initial density profile is $\rho=T^m$
  and the initial pressure profile is $P = T^{m+1}$. Our initial conditions are in hydrostatic
  and thermal equilibrium, but are superadiabatically-stratified and unstable to convection. As shown
  in Figure \ref{setup}, the polytropic temperature-pressure profile does a good job of approximating the
  realistic profile for a directly imaged giant planet used in \citet{Moses2016}.
  
\subsection{Reactive tracer field}
  To this system, we add a reactive tracer $c$ that reacts according to the reaction,
  \begin{align}\label{crxn}
    \ce{X2 + c ->[k] Y + Z},
  \end{align}
  where X$_2$ is presumed to be the dominant component of the atmosphere
  (i.e. of density approximately equal to $\rho_0$, the background density), and k is the forward
  reaction rate constant, constructed as an unmodified Arrhenius rate law,
  \begin{equation}
    \text{k} = \frac{1}{\tau(z=0)\rho_0}e^{-T_\text{act}/T_0},
  \end{equation}
  where $\tau(z=0)$ is a fixed value for the timescale of the reaction at the
  bottom of the atmosphere, with $T_\text{act}$ the fixed activation temperature
  for the reaction. The choice of these two values sets the ratio of chemical
  to density scale heights and the vertical profile of the chemical timescale.
  %This rate law is defined in terms of background thermodynamic quantities
  %as variations in the full quantities $\rho$ and $T$ will be of order $\epsilon$,
  %and therefore not significantly impact the chemical reaction rate.
  The reactive tracer
  %\footnote{ Is it worth mentioning anything about the passive tracer?}
  is evolved along with the dynamics,
  \begin{equation}
    \frac{\partial c}{\partial t} + \mathbf{u}\pmb{\cdot\nabla} c = 
    \frac{1}{\rho}\pmb{\nabla\cdot} \left[D\rho\pmb{\nabla} c\right] + \text{k}\rho_0(c_\text{eq} - c) ,
  \end{equation}
  where $c$ is the mole fraction of our reactive tracer, $\rho_0$ is the background
  density profile, and $c_\text{eq}$ is the equilibrium profile of $c$. $D$ is the molecular diffusivity,
  set by the Schmidt number, which for simplicity's sake we set to one
  (i.e., $\nu = D$). Background profiles are used for the concentration of X$_2$,
  and the Arrhenius rate law due to the fact that
  fluctuations in the density variable will be of order $\epsilon$ \citep{Anders2017}.
  
  We chose to work with two different equilibrium profiles in this work, as shown in
  the right panel of Figure \ref{setup}, aimed at bracketing the two extremes of equilibrium profiles
  relevant to real chemical species. Specifically, those 
  that vary slowly with height (i.e. approximately linear or constant profiles)
  and those that vary rapidly with height (i.e. profiles that can be approximated using
  tanh functions). By including two different choices of equilibrium profiles, we are able
  to explore species-dependent effects on quenching behavior.
  
%  \begin{table*}[!ht]
%    \centering
%    \begin{splittabular}{cccBccc}
%      \hline & 2-D & & & 3-D &\\ \hline
%      log$_{10}Ra$ & Resolution (Z $\times$ X) & Evolution time ($\tau_B$) &
%      log$_{10}Ra$ & Resolution (Z $\times$ X $\times$ Y) & Evolution time ($\tau_B$)
%      \\ \hline\hline
%      4 & 128 $\times$ 512 & 500 & 4 & 128 $\times$ 256$^2$ & 480\\
%      4.25 & 128 $\times$ 512 & 1420 &  &  & \\
%      4.5 & 192 $\times$ 768 & 1140 & 4.5 & 128 $\times$ 256$^2$ & 740\\
%      4.75 & 192 $\times$ 768 & 1230 &  &  & \\
%      5 & 256 $\times$ 1024 & 980 & 5 & 256 $\times$ 512$^2$ & 460\\
%      5.25 & 256 $\times$ 1024 & 820 &  &  & \\
%      5.5 & 256 $\times$ 1024 & 830 & 5.5 & 256 $\times$ 512$^2$ & 610\\
%      5.75 & 384 $\times$ 1536 & 500 &  &  & \\
%      6 & 512 $\times$ 2048 & 1390 & 6 & 256 $\times$ 512$^2$ & 1150\\
%      6.25 & 768 $\times$ 3072 & 1680 &  &  & \\
%      6.5 & 768 $\times$ 3072 & 1750 &  &  & \\
%      6.75 & 768 $\times$ 3072 & 2120 &  &  & \\
%     7 & 1024$\times$2048 & 2430 & & & \\ \hline
%    \end{splittabular}
%    \caption{\label{simlist}List of simulations. All simulations are performed
%      with a vertical stratification of three density scale heights and an aspect
%      ratio of four. Evolution times are given in units of buoyancy times ($\tau_B$).}
%  \end{table}
    \begin{table}[!ht]
    \centering
    \begin{tabular}{ccc}
      \hline & 2-D &\\ \hline
      log$_{10}Ra$ & Resolution (Z $\times$ X) & Evolution time ($\tau_B$) \\
      \hline\hline
      4 & 128 $\times$ 512 & 500 \\
      4.25 & 128 $\times$ 512 & 1420 \\
      4.5 & 192 $\times$ 768 & 1140 \\
      4.75 & 192 $\times$ 768 & 1230 \\
      5 & 256 $\times$ 1024 & 980 \\
      5.25 & 256 $\times$ 1024 & 820 \\
      5.5 & 256 $\times$ 1024 & 830 \\
      5.75 & 384 $\times$ 1536 & 500 \\
      6 & 512 $\times$ 2048 & 1390 \\
      6.25 & 768 $\times$ 3072 & 1680 \\
      6.5 & 768 $\times$ 3072 & 1750 \\
      6.75 & 768 $\times$ 3072 & 2120 \\
      7 & 1024$\times$2048 & 2430\\\hline
      & 3-D & \\ \hline
      log$_{10}Ra$ & Resolution (Z $\times$ X $\times$ Y) & Evolution time ($\tau_B$)\\\hline\hline
      4 & 128 $\times$ 256$^2$ & 480\\
      4.5 & 128 $\times$ 256$^2$ & 740\\
      5 & 256 $\times$ 512$^2$ & 460\\
      5.5 & 256 $\times$ 512$^2$ & 610\\
      6 & 256 $\times$ 512$^2$ & 1150\\\hline
    \end{tabular}
    \caption{\label{simlist}List of simulations. Shown are Rayleigh numbers $Ra$,
      coefficient resolutions, and evolution times in buoyancy time units, $\tau_B$ (see text).
      All simulations are performed
      with an initial vertical stratification of three density scale heights and an aspect
      ratio of four.}
  \end{table}
  \subsection{Numerics}
  In this work, we utilize the open-source framework Dedalus\footnote{Dedalus is available at
  \url{http://dedalus-project.org}} \citep{Burns2017}, which has been successfully
  used to study atmospheric dynamics
  \cite[e.g.,][]{Lecoanet2014,Lecoanet2015_water,Lecoanet2016,Lecoanet2016_KH,Anders2017}.
  Dedalus is a pseudo-spectral code that employs implicit-explicit timestepping and solves linear terms
  implicitly in spectral space, and nonlinear terms explicitly in physical space \citep{Burns2016,Burns2017}.
  Our simulations resolve the physical viscosity of the simulation on the grid, obviating the need for
  artificial hyperdiffusive terms, or filters that suppress artificial modes in the numerical solution
  \citep[e.g.,][]{Cooper2006}. In our simulations, all field variables are
  represented with a Fourier basis in
  the horizontal, and a Chebyshev basis in the vertical {\citep{Boyd2001}}. The code evaluates nonlinear terms on a grid
  with a factor 3/2 more points than Fourier coefficients. Resolutions are reported in Table \ref{simlist}.

  In our simulations, the horizontal boundaries are periodic. On the upper and lower
  boundaries we employ stress-free, impenetrable, fixed thermal flux, and zero tracer
  flux boundary conditions. We initiate our simulations by setting the fluctuating
  temperature component field to be a noise field of small perturbations
  (of order 0.01$\epsilon$ or smaller). Initially, the fluctuating component of
  density is set equal to zero, and the tracer fields begin in equilibrium.
  We have performed experiments using different sets of initial conditions for the tracer fields and
    found no dependence of the tracers' evolution on their initial
    conditions.

  \section{Results}\label{sec:results}
  \subsection{Dynamic evolution}\label{sec:dyresults}
\begin{figure}
  \centering
  \includegraphics[width=3.5in]{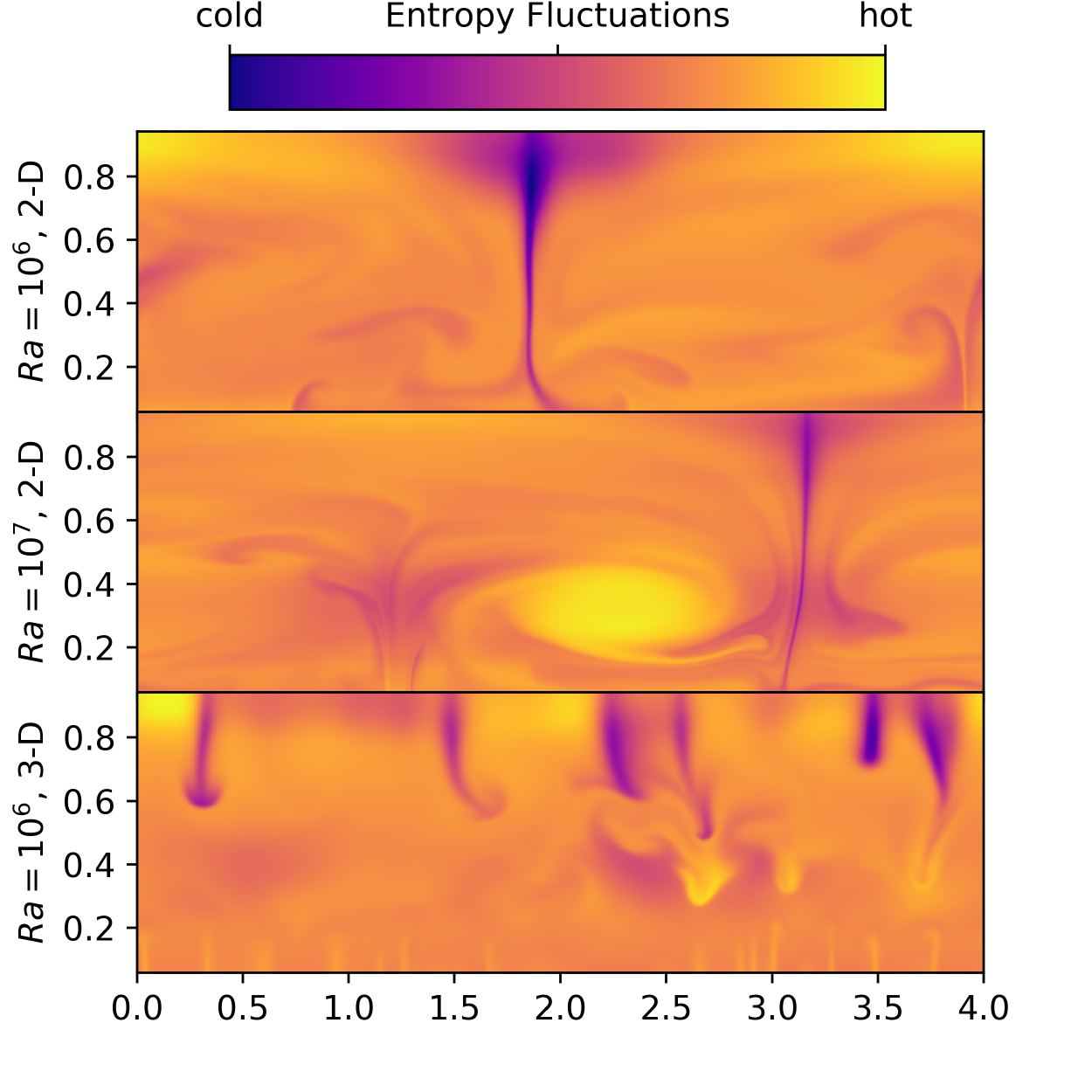}
  \caption{Snapshots of the dynamic evolution of 2-D Rayleigh number runs
    at 1200 buoyancy times, and a 3-D case at 290. The top case demonstrates
    the primarily laminar flow
    regime that characterizes the lower $Ra$ end of our parameter space. In the
    middle, the result of several large scale structure mergers dominates the
    mixing in the
    second dynamic regime. At the bottom, a slice through the center of a 3-D
    simulation shows a plume-dominated structure.
    \label{slices}}
\end{figure}

  All simulations are dynamically evolved to a steady state characterized by a
  relatively constant rms vertical velocity, with evolution times described in Table
  \ref{simlist}. With increasing Rayleigh number, our 2-D simulations transitioned from
  laminar flows which settled on a constant rms vertical velocity quickly and
  showed no long term evolution ($  10^4 \le Ra < 10^6$) to flows initially dominated
  by multiple large-scale circulating structures which eventually underwent merger events prior to very slowly
  evolving towards a constant rms vertical velocity ($10^6 < Ra \le 10^7$). %Based on
  %previous work, we expect that we will find a third dynamic regime as we approach our
  %maximum simulated $Ra$.
  In contrast, in 3-D simulations, plume structures at the top and bottom of the domain
  dominate the transport, and large-scale circulating structures never develop due to the lack of a
  preferred axis.
  These three dynamic regimes are showcased in Figure \ref{slices},
  which compares two different 2-D $Ra$ cases captured at the same time in their dynamical
  evolution, and a 3-D case that has reached a similarly evolved state after a shorter evolutionary
  period.

The demand for higher resolution and substantial long term evolution of the higher $Ra$
2-D cases only further highlights the significant computational
challenges of accessing realistic regimes for planetary atmospheres \citep{Showman2011}.
This is due to the establishment of the large-scale circulating structures mentioned above that
span the domain in 2-D (which are not present in 3-D).
In 3-D, shorter evolution times, due to the lack of these slowly evolving, circulating structures
make runs less challenging. The increased resolution requirements
due to the extra dimension, however, adds significantly to computational cost. Given
that realistic $Ra$ are inaccessible to atmospheric modeling, we
explore the relationship between $Ra$ and the mean transport properties of the atmosphere
while acknowledging the caveat that further dynamic regimes may exist that do not scale
with what we observe and infer here.
    
\subsection{Disequilibrium chemistry}
\begin{figure}
      \centering
      \includegraphics[width=3.5in]{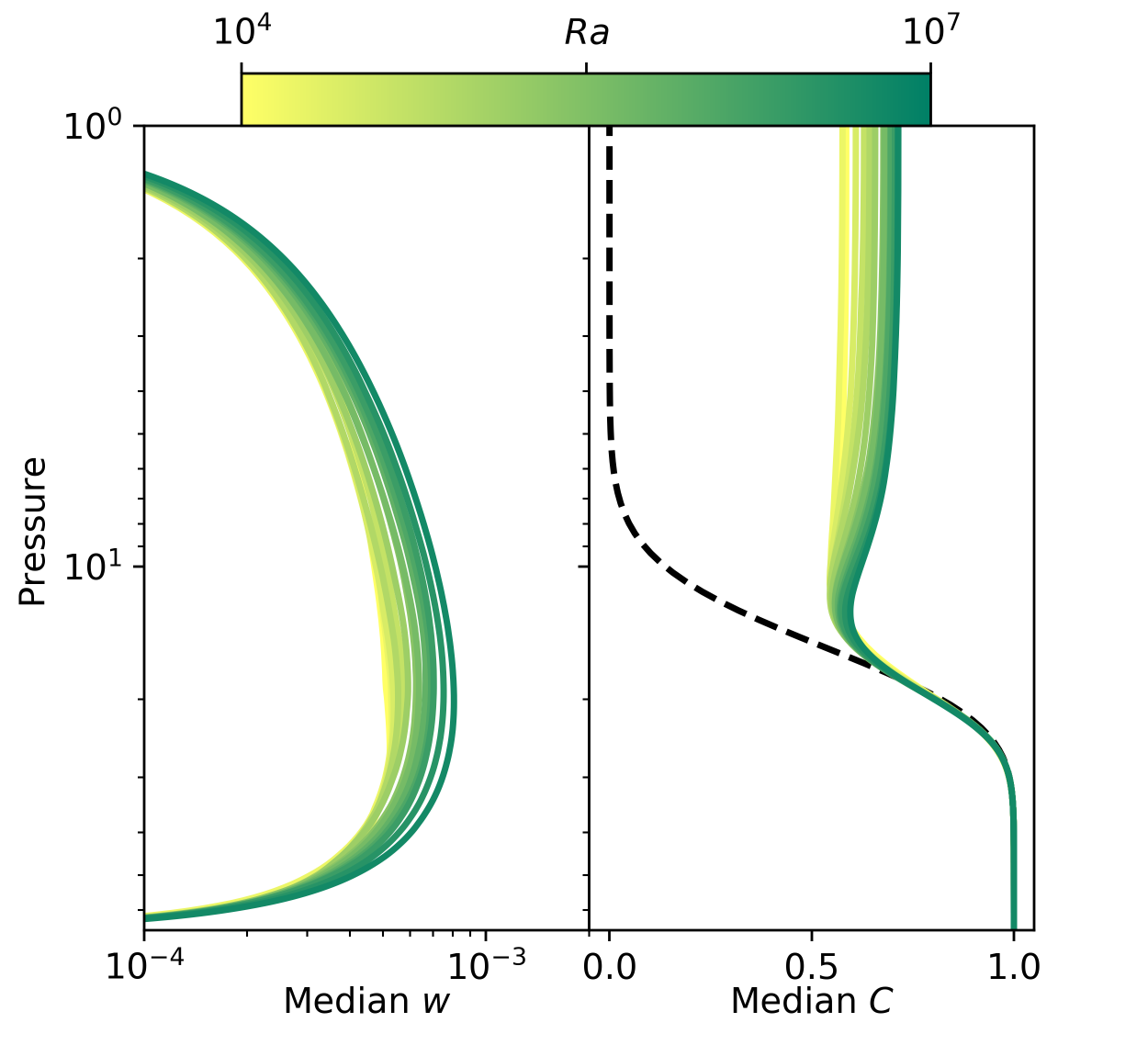}
      \caption{Horizontally averaged vertical profiles of the vertical velocity and
        reactive tracer mole fraction, medianed over the last 200 buoyancy times of the
        2-D tracer evolution.
        In the right panel, the dashed black line is the tanh equilibrium profile used
        in the evolution of the reactive tracer for this set of simulations.
        As the median velocity of the dynamics
        increases with increasing Rayleigh number, the quench point inferred from the
        top of the simulation moves deeper into the atmosphere. \label{vert_profiles}}
\end{figure}
\begin{figure}
      \centering
      \includegraphics[width=3.5in]{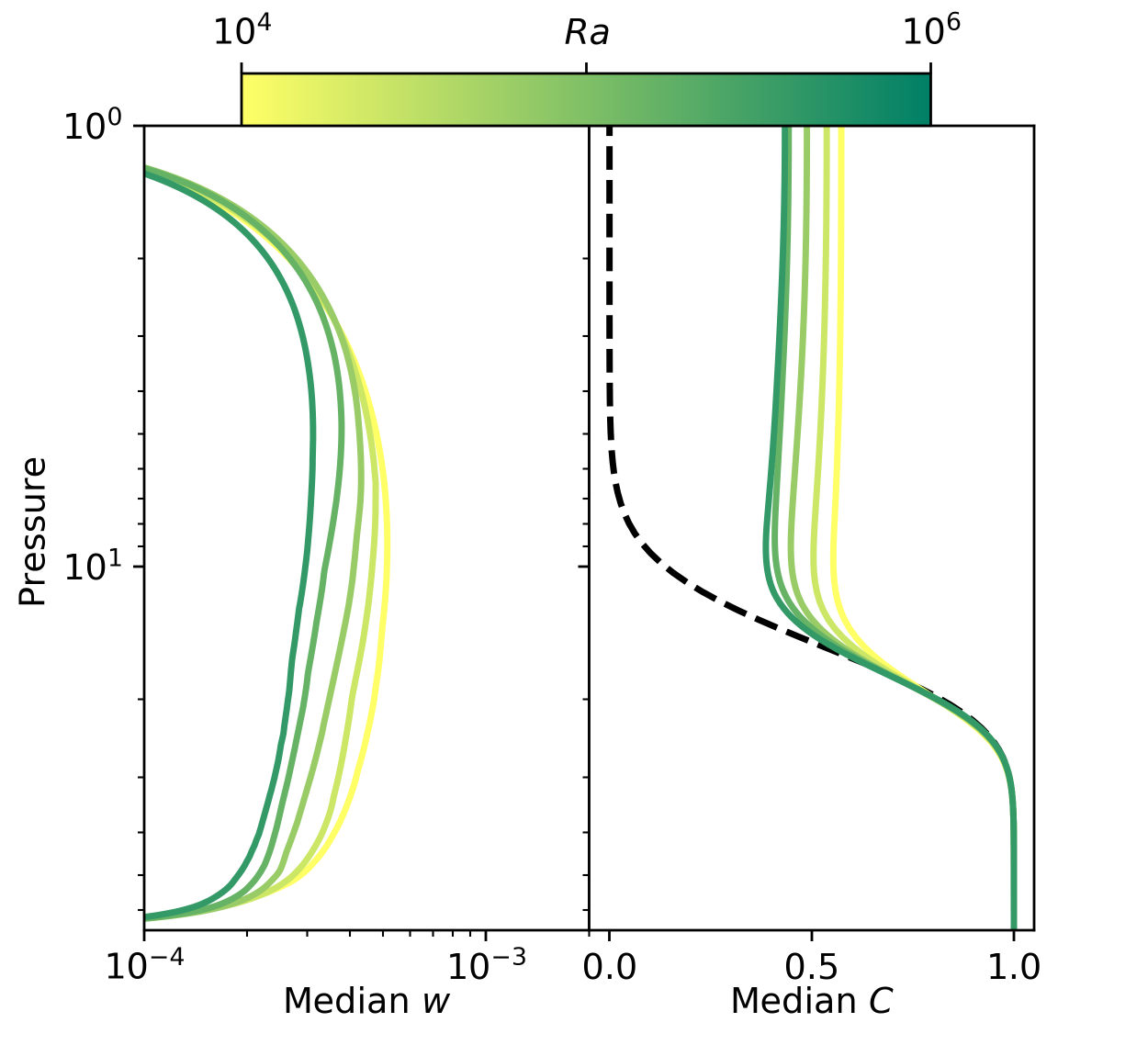}
      \caption{The same quantities shown in Figure \ref{vert_profiles} for the 3-D cases.
        The tracer profiles are medianed over between 25 and 150 buoyancy times after they
        reach a steady state. These cases were not evolved for as long as the 2-D cases,
        as they reached a steady state more quickly and with significantly less variation.
        In the 3-D cases, the median vertical velocity of the dynamics decreases with increasing
        Rayleigh number, and the quench point inferred from the top of the simulation moves higher
        in the atmosphere.
        \label{vert_profiles3D}}
\end{figure}

\begin{figure*}
      \centering
      \includegraphics[width=7in]{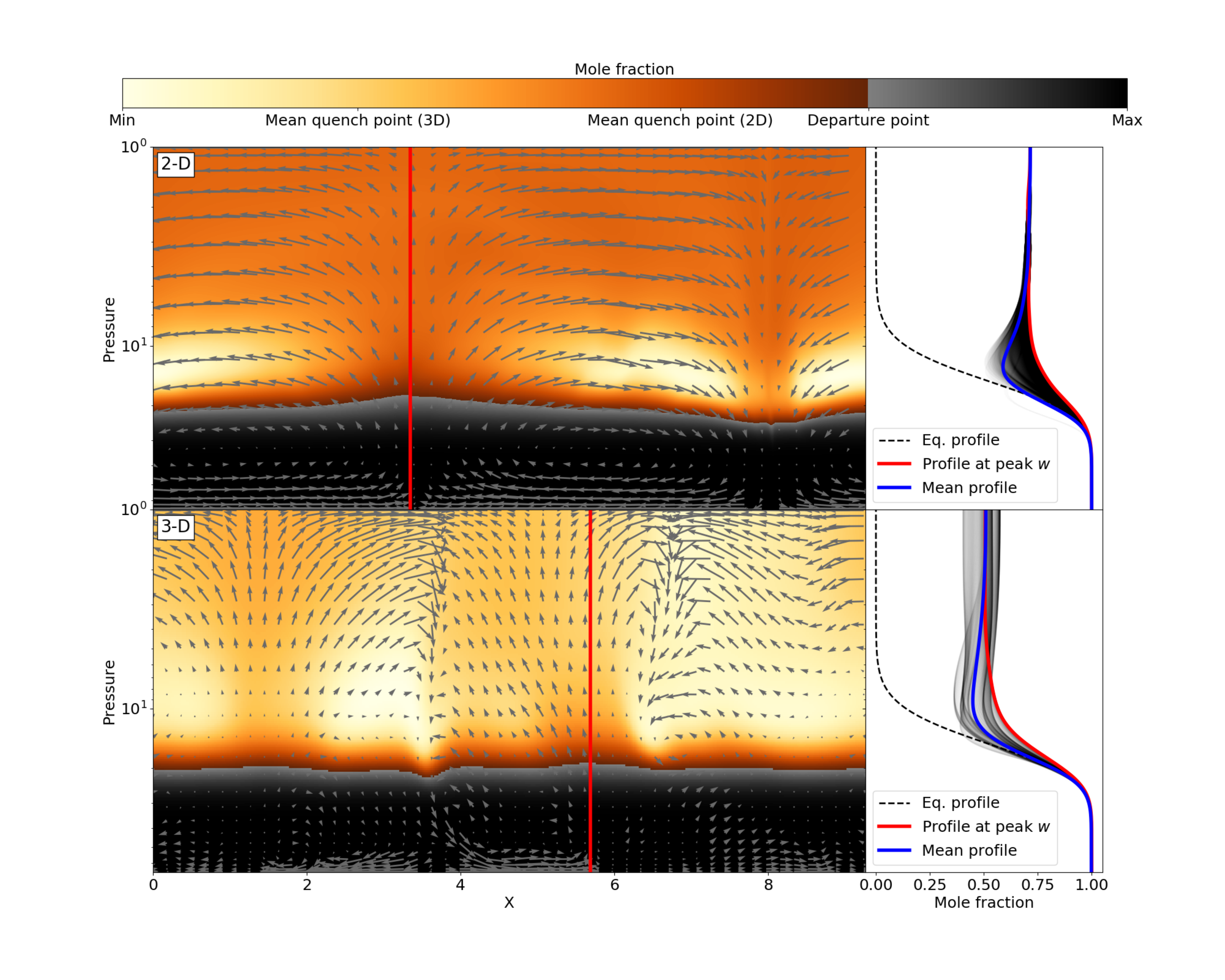}
      \caption{{Left: 2-D slices of the reactive tracer distribution in steady state for the 2-D $Ra=10^7$
        (top)
        and 3-D $Ra=10^5$ (bottom) simulations. The 3-D slice is taken from halfway through the y-axis
        of the
        domain. Arrows indicate the direction and magnitude of the flow field
        within the slice. The location of the peak vertical velocity $w$ is indicated with a red
        vertical line. Right: The corresponding profiles at each horizontal position, with opacity
        weighted by the strength of the vertical velocity at that position. The red profile corresponds
        to the horizontal position with the peak vertical velocity, while the blue line describes the
        mean profile across all horizontal positions. The overall quench point is determined by the
        locations of the strongest vertical flows and propagated horizontally by strong horizontal flows
        in the upper region of the domain (especially prominent in 2-D),
        while the departure point and transition region are
        significantly affected by local flow structures in both 2- and 3-D.}
        \label{2D3Dflowplots}}
\end{figure*}

    When each simulation was evolved to the dynamically steady state described  in Section
    \ref{sec:dyresults}, reactive tracers were injected, and the simulations were evolved
    forward a further five hundred buoyancy times in the 2-D cases, and roughly one hundred
    buoyancy times in the 3-D cases, where, $\tau_B = \sqrt{L_z/g\epsilon}$.
    In all 2-D cases, these reactive tracers
    were observed to reach a steady state, as measured by a lack of evolution in various
    measures of the quench point, within approximately fifty buoyancy times. In the 3-D cases,
    this convergence happened even more rapidly, taking approximately thirty buoyancy times.

    Two different metrics are used to measure the quenching behavior of the reactive tracers in
    our simulations. The first metric is an ``observed'' quench point, where the mole fraction of
    the reactive tracer at the top of the atmosphere is used as a proxy for what would be
    observed in an atmosphere remotely. The corresponding pressure for where the equilibrium
    profile of the reactive species takes on that value is taken to be the quench point,
    making the standard assumption that the quenching species immediately takes on the
    equilibrium value where it begins to quench. The second metric is a measured departure
    point, which is the location at which the reactive species profile moves above the
    equilibrium profile. This metric allows for comparison of the value at which the tracer
    is observed to quench to the location at which the profile actually begins to diverge
    from the equilibrium profile.

    Figures \ref{vert_profiles} and \ref{vert_profiles3D} show the 2-D and 3-D median vertical
    profiles of vertical velocity and
    the mole fraction of the reactive tracer, averaged horizontally and medianed {in time} over
    the end portions of their evolution. In 2-D, with increasing $Ra$, the vertical
    velocity profile shifts to slightly higher values (left panel), leading to an observed quench
    point deeper in the atmosphere (right panel). As the vertical velocity profiles increase in
    magnitude, the quench point inferred from the top of the atmosphere moves deeper into
    the atmosphere. In contrast, in 3-D, the vertical velocity profiles decrease in magnitude
    with increasing $Ra$, and the observed quench point rises in the atmosphere. In the 2-D case,
    there is also a pronounced transition
    region between the departure point and the height in the atmosphere at
    which the reactive tracer profile takes on a constant value. {
      This behavior, which is reduced in the
    3-D case, is likely an effect of the constrained 2-D dynamics described in Section
    \ref{sec:dyresults}. This is demonstrated in Figure \ref{2D3Dflowplots}, which shows the
    two-dimensional  distribution of the reactive tracer and the dynamical flow patterns. In the 2-D
    case (top), the large-scale circulating flow structures trap fluid in their relatively stagnant
    centers (bright regions), leading to regions without significant quenching and therefore lower
    concentrations of the reactive tracer. Above these stagnant centers, horizontal flows homogenize
    the upper levels of the profile to the value determined by the region of strongest vertical flow.
    In the 3-D case (bottom), these large-scale circulating structures are weaker and more local,  and
    so strong vertical flows mix the fluid to different quench values that are not smoothed out across
    the entire domain as is true in the 2-D case. The region of strongest vertical flow, however, still
    dominates in setting the quench point determined by the mean profile.}

\begin{figure}[!ht]
      \centering
      \includegraphics[width=3.5in]{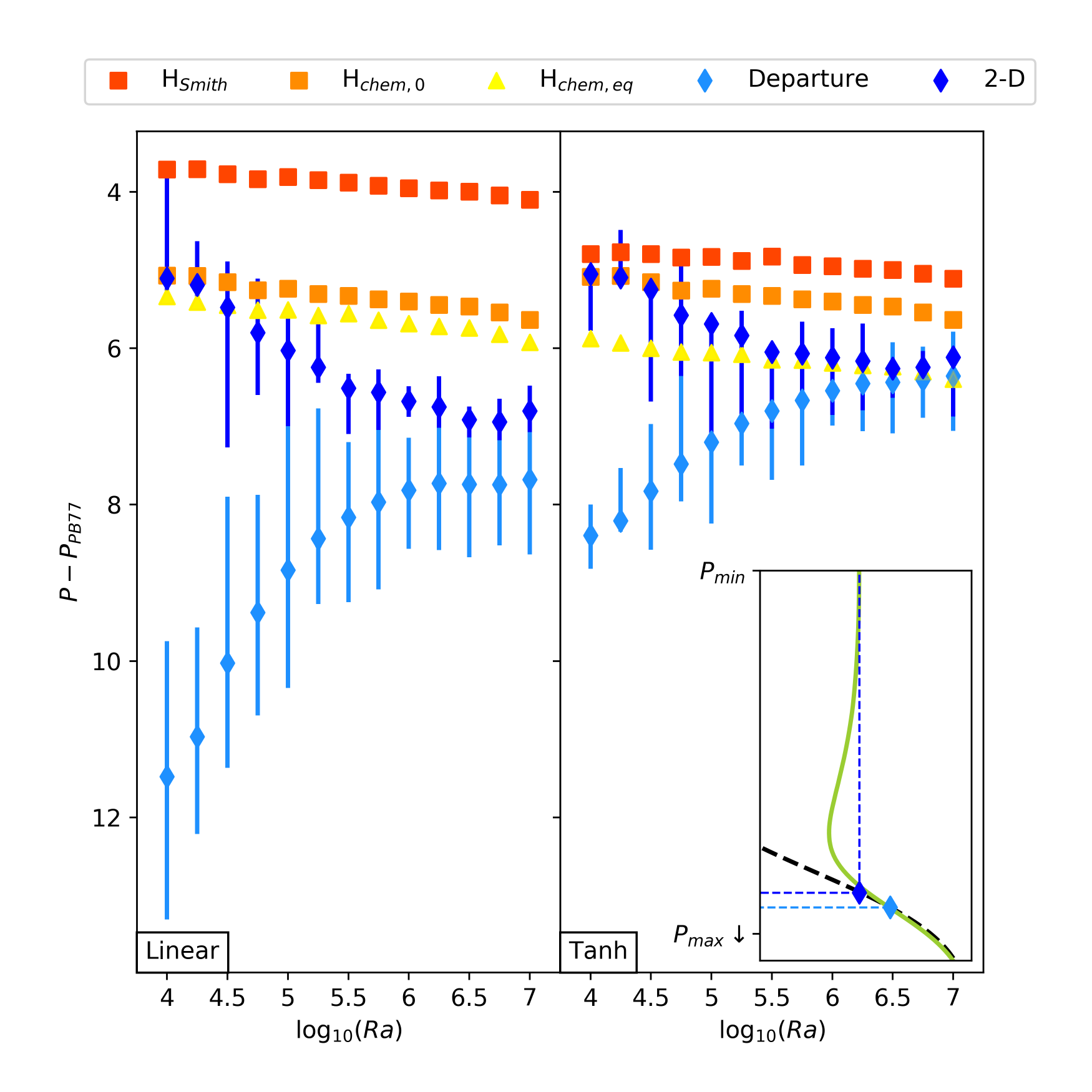}
      \caption{The deviation of various measured and predicted quench points from the
        classical prediction of PB77 for the linear (left) and tanh (right)
        equilibrium profile cases for the 2-D simulations. The square markers indicate predicted quench points
        based on the Smith length scale and the chemical scale height in Equation \ref{chemical_scaleheight_0}.
        The triangular markers
        are predicted quench points using the chemical scale height in Equation \ref{chemical_scaleheight_eq}.
        The diamond markers are measured quench points,
        where the departure points are the pressure at which the median profile of the
        reactive tracer moves above its equilibrium profile, and the 2-D and 1-D points
        are the corresponding pressures to the concentration measured at the top of the median
        tracer profile. Error bars indicate the maximum deviation of these measured quench
        points from their median value once they have reached a steady state. 
        \label{QP}}
    \end{figure}

\begin{figure}[!ht]
      \centering
      \includegraphics[width=3.5in]{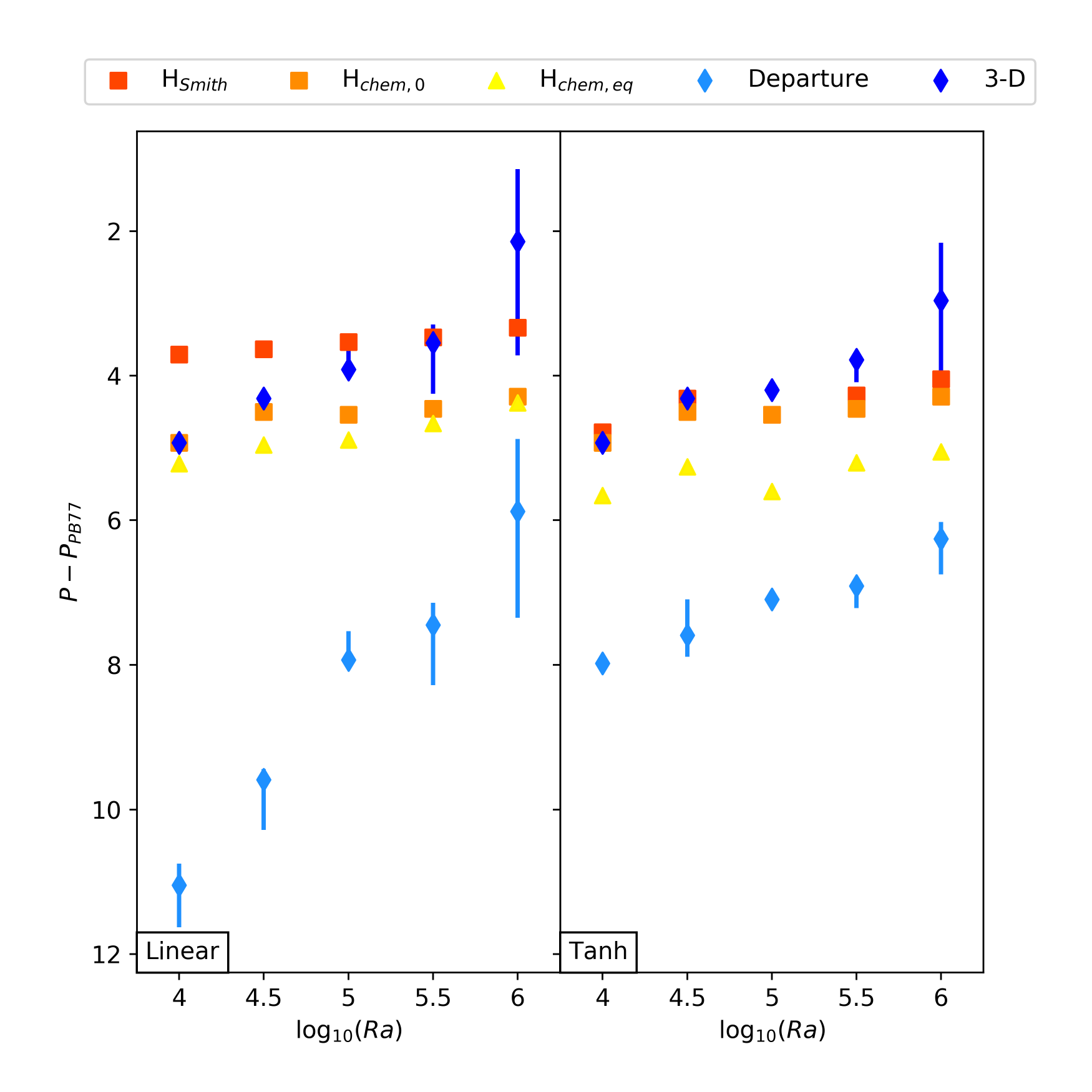}
      \caption{The same measurements made in Figure \ref{QP} for the 3-D simulations.
        \label{QP3D}}
    \end{figure}

    Figures \ref{QP} and \ref{QP3D} shows various measured and predicted quench points relative to
    the classical quench point predicted using the theory described by PB77.
    Three salient trends emerge in Figure \ref{QP} for both equilibrium profile cases in 2-D:
    1) significantly deeper quenching than predicted by PB77, 2) the steady
    approach of the departure point towards some depth intermediate between the
    classic PB77 prediction and the departure measured at the lowest $Ra$
    probed, and 3) the slow convergence between the departure point and the measured
    quench point.

    Based on these trends, some results of the quench approximation
    may reasonably be expected to hold at realistic $Ra$. Namely, that the quenched profile will be
    rapidly achieved, such that the measured quench point will accurately track the
    point at which the equilibrium and actual chemical profiles diverge.
    Comparing the two cases, the slowly varying linear profile clearly has a
    wider spread of predicted and measured quench points than the rapidly varying tanh
    profile. This speaks to the fact that for species where quenching occurs in a relatively
    constant region of their equilibrium profile, measuring and predicting quench points
    is necessarily more difficult due to the fact that a very small difference in mole fraction
    may reflect a large change in pressure.

    In our 3-D simulations, trends are present, but not as clearly as in our 2-D cases.
    Both the measured quench point and
    the departure point continue the trend shown in 2-D of quenching deeper in the atmosphere than
    predicted by PB77, but whether or not the departure point is converging towards a point intermediate
    between that measured at the lowest $Ra$ and the prediction of PB77, and towards the measured
    quench point is less clear. Although the measured quench point and the departure point may converge
    at much higher $Ra$, it is not clear from the $Ra$ probed that this is the case, making it difficult
    to predict from the quench point where the chemical profile actually diverges from the equilibrium
    profile. Similarly to the 2-D cases, however, we see that the tanh profile has a narrower range
    of measured quench depths and departure points, indicating that our interpretation from our
    2-D results, that it is easier to measure quenching in a rapidly varying region of a chemical profile,
    holds in 3-D.

    As was described in \cite{Smith1998} and confirmed in \citet{Cooper2006}, a
    significantly smaller characteristic length scale, $H_\text{Smith}$,
    than the density scale height is
    required to match our results to predictions made using mixing length theory. 
    We find, however, that the Smith length scale is still too large to accurately
    predict the behavior of our reactive tracer in 2-D.

    A slightly better metric is
    the pure chemical length scale described in PB77, which in our formulation
    takes the form,
    \begin{equation}\label{full_chemical_scaleheight}
      \begin{split}
      \text{H}_\text{chem} &= \left[-\frac{d}{dz}\ln[k\rho^2c]\right]^{-1}\\
      &= \left[\frac{1}{H_k}+\frac{2}{H_\rho}-\frac{d}{dz}\ln(c)\right]^{-1},
      %&= \left[\frac{1}{H_k}+\frac{1}{H_\rho}-\frac{d}{dz}\ln(c_\text{eq}-c)\right]^{-1}
      \end{split}
    \end{equation}
    where $H_k$ and $H_\rho$ are the scale heights of the rate constant and density respectively.
    {A
    general derivation of this length scale is given in the Appendix.}
    This length scale is simpler to derive than the Smith length scale
    in the limit that the departure point accurately predicts the observed quench point (i.e.
    if $c$ varies very slowly with height after quenching), such that the third term can
    be taken to be zero,
    \begin{equation}\label{chemical_scaleheight_0}
      \text{H}_\text{chem,0} = \left[\frac{1}{H_k}+\frac{2}{H_\rho}\right]^{-1}.
    \end{equation}
    An even better estimation, as shown in Figure \ref{QP} as $\text{H}_\text{chem,eq}$,
    involves approximating the third term by setting $c$ equal to its equilibrium profile (as should
    be valid at the quench point),
    \begin{equation}\label{chemical_scaleheight_eq}
      \text{H}_\text{chem,eq} = \left[\frac{1}{H_k}+\frac{2}{H_\rho}-\frac{d}{dz}\ln(c_\text{eq})\right]^{-1},
    \end{equation}
    which is only slightly more difficult to calculate from known chemical data. 

    In 3-D, it is not clear from our results that one
    length scale does significantly better than another in predicting the measured quench point,
    as it crosses the predictions of the various length scales with increasing $Ra$.
    In this case, H$_\text{chem,eq}$ seems
    to still do the best job of predicting the departure point, but it is difficult to infer if
    this will continue to be the case with increased $Ra$.
    %Moreover, as is shown in Figure \ref{QP}, the deviation of the measured quench point
    %from the predictions of mixing length theory increase with increasing $Ra$, implying
    %a relationship between this
    %lengthscale and the underlying dynamics.
    
    \subsection{1-D Chemical Modeling}\label{sec:chemmodel}
    \begin{figure}[!ht]
      \centering
      \includegraphics[width=3.5in]{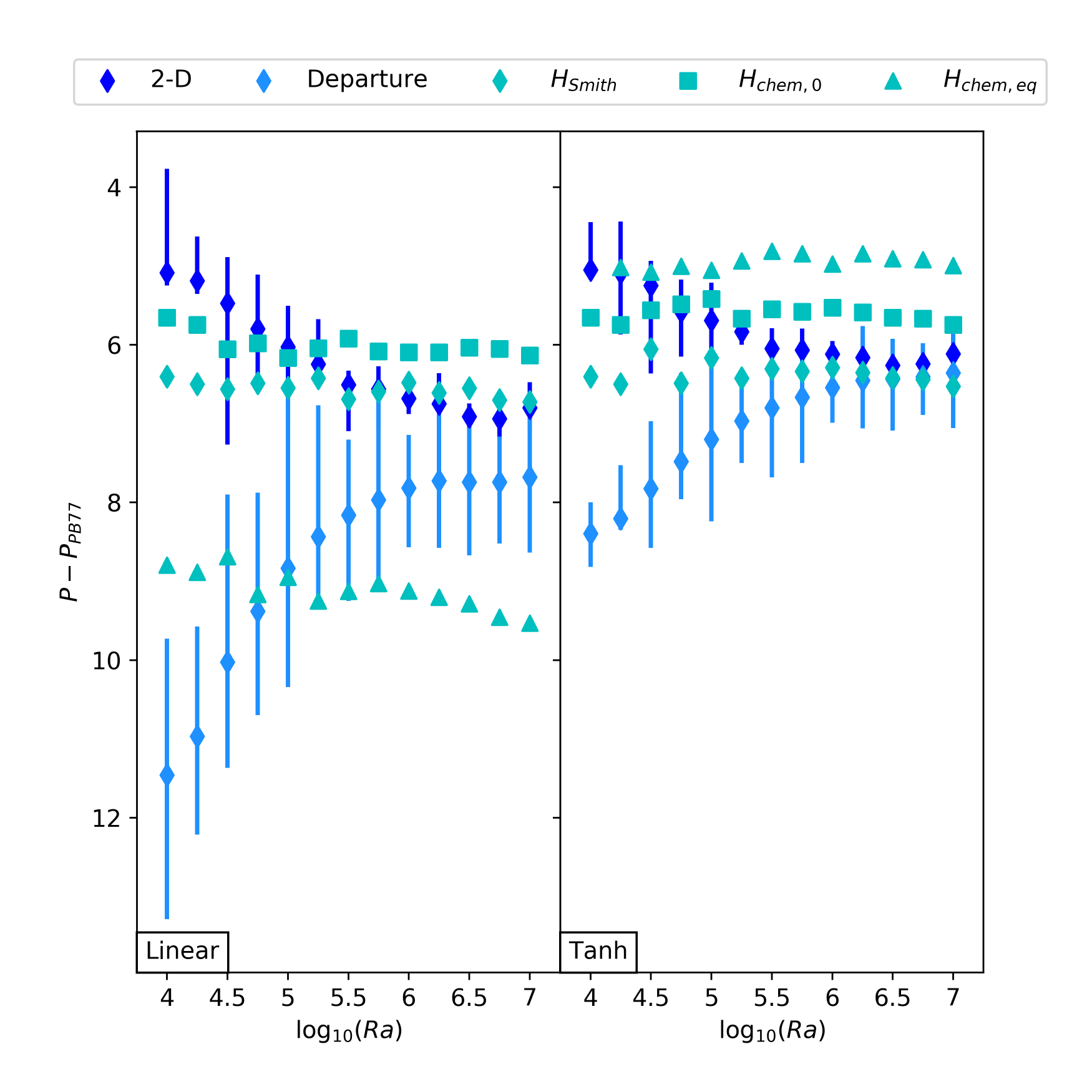}
      \caption{The results of the 1-D models using different eddy diffusion coefficients for each of
        the three length scales under consideration compared to the 2-D simulations. The teal points represent 1-D results using the various
        ascribed length scales.
        \label{QP1D}}
    \end{figure}

    \begin{figure}[!ht]
      \centering
      \includegraphics[width=3.5in]{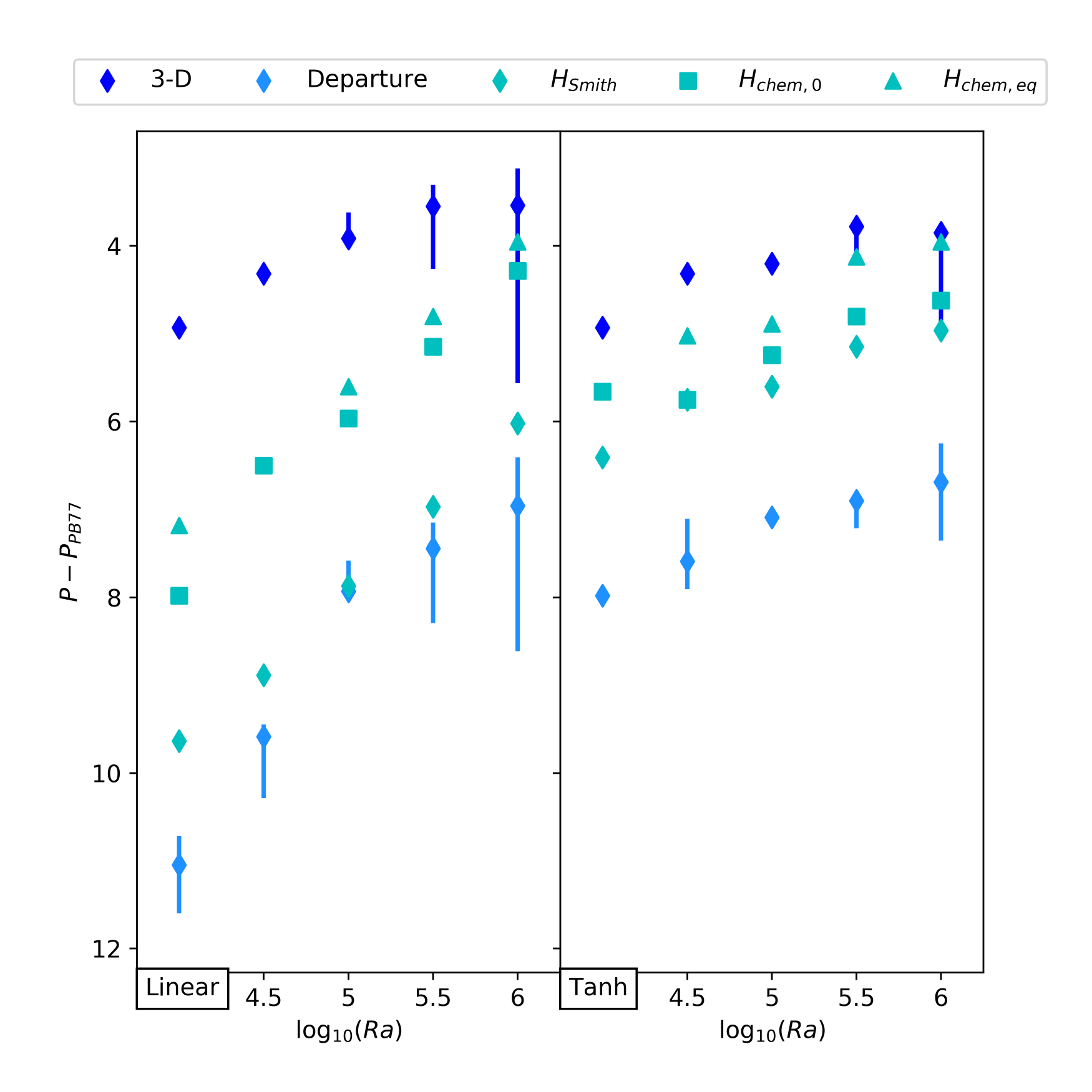}
      \caption{The same measurements made in Figure \ref{QP1D} for
        the 3-D simulations.
        \label{QP1D3D}}
    \end{figure}

    In addition to our 2- and 3-D models, we created a 1-D eddy diffusion model that
    solves for the equilibrium profile of the reactive tracer according to,
    \begin{equation}
      \frac{\partial c}{\partial t} + w\frac{\partial c}{\partial z} =
      \frac{1}{\rho_0}\frac{\partial}{\partial z}\left[(D+K_\text{zz})\rho\frac{\partial c}{\partial z}\right] +\text{k}\rho_0(c_\text{eq} - c),
    \end{equation}
    where $w$ is the median profile of the vertical velocity calculated over the course of the last
    five hundred buoyancy times of our simulations' evolution, and the eddy diffusion is calculated
    using that same velocity profile and the various length scales probed in Figures \ref{QP} and
    \ref{QP3D}
    following Equation \ref{eddy_diff}. These models are solved as a nonlinear boundary value problem
    with $\partial/\partial t = 0$.
    
    In 2-D and 3-D, as shown in Figures \ref{QP1D} and \ref{QP1D3D}, 1-D models using eddy
    diffusion coefficients involving the chemical equilibrium profile and rate law (e.g., $H_\text{Smith}$ or $H_\text{chem}$) do a relatively
    good job of modeling the quenching behavior we measured from our simulations. No one
    eddy diffusion model (based on the different length scales) stands out strongly as the preferred
    model to use for 1-D chemical modeling. In 2-D, there is some spread in the results of the
    different models, though the Smith length scale seems to do marginally better at predicting the
    observed quench point and departure point. In 3-D, there is also some spread to the results, but the best model to use is not clear.
    The fact that in the case of the 1-D models no one length scale wins out speaks
    to the difference between modeling small scale dynamics where
    the disequilibrium chemistry is occurring at, and the dynamics driving transport at
    large scales.

    %The fact that the Smith lengthscale
    %produces 1-D results that accurately predict the 2-D \textbf{(and 3-D?)} results,
    %but fails when used to predict the 2-D results based on timescale arguments is
    %due to the fact that \citet{Smith1998} constructs his lengthscale based on matching
    %his explicit diffusion model to a diffusive mixing model (analogous to our 1-D
    %model)\footnote{This point needs to be clarified, ideas? I'm basing this claim on his Eqn 19,
    %which holds velocity (mixing) constant, rather than the dynamical quench point.}

    %\begin{figure}
    %  \centering
    %  \includegraphics[width=3.5in]{quenchin2.pdf}
    %  \caption{The deviation of various measured and predicted quench points from the
    %    departure point from the equilibrium curve for the linear (top) and tanh (bottom)
    %    equilibrium profiles.
    %    \label{QP2}}
    %\end{figure}

\vspace{1cm}
\section{Conclusion} \label{sec:conclusion}
In this work we take a different tack from previous dynamical studies exploring the
transport properties of the atmospheres of weakly irradiated
giant planets and brown dwarfs, and examine the relationship between chemical
disequilibrium behavior and the dynamic regime being simulated. We parameterize the dynamic
driving by the Rayleigh number, $Ra$. Using 2- and 3-D local box models of vertically stratified
convective regions in a parameter space relevant to these objects, we 
have confirmed that the simple assumption of a chemically-independent characteristic
length scale is not sufficient to explain the quenching behavior of a reactive tracer.
Instead the length scale depends on the chemical properties of the reacting species
(as was demonstrated for the 1-D case by \citet{Smith1998}). Further, we have shown
that there exists another length scale, the chemical scale height, $H_\text{chem,eq}$ (Equation
\ref{chemical_scaleheight_eq}), which more accurately
predicts the measured quench point, can be incorporated effectively into an eddy
diffusion coefficient for 1-D chemical kinetics modeling,
and is simpler to calculate than the Smith length scale.
Finally, we have shown that for the $Ra$ that are currently accessible to numerical simulation,
the transport properties of the atmosphere depend on the dynamic regime being simulated.
This is an important consideration as the field moves forward towards coupling complex
chemistry more directly with more realistic atmospheric dynamics
\citep{Heng2015,Drummond2016}.

In this paper we have focused on validating the quench approximation in the convective
regions of substellar atmospheres. Overlying these convective regions, however, there
are the stably stratified radiative zones. In these zones, the basis of the quench approximation,
the eddy diffusion approximation, does not necessarily hold, and the effect of the transport
properties of these zones upon quenching species may be very different \citep{Parmentier2013}.
Additionally, even if the eddy diffusion approximation does reasonably describe the
transport properties of the radiative zone, for species that quench within these regions,
differences in the structure of the temperature-pressure profile may lead to quenching regimes where
current models for estimating the eddy diffusivity no longer apply (such as regions where the
chemical timescale increases very slowly). %\footnote{I need to
%  think on this claim a tiny bit more, as partial pressures may come into play in a way that
%  negates the effect of an isothermal atmosphere...I don't think so?}.
We look forward to
further investigation into the dynamics of these regions, and their interaction with the
chemistry of the atmospheres of giant planets and brown dwarfs.

\acknowledgements
Computations were conducted, with support from the NASA High End Computing (HEC) Program through
the
NASA Advanced Supercomputing (NAS) Division at Ames Research Center, on Pleiades with allocations
GID s1647 (PI: Brown) and GID s1419 (PI: Toomre). Financial support was provided by University of Colorado Boulder and Bates College startup funding. Additionally, this work benefited from the
Exoplanet Summer Program in the Other Worlds Laboratory (OWL) at the
University of California, Santa Cruz, a program funded by the Heising-Simons Foundation.
The authors would like to thank Mark Rast, Julianne Moses, and Vivien Parmentier for
useful discussions. We thank the anonymous referee for useful comments that improved the quality of the paper.

{\software{Dedalus \citep{Burns2017}}}
\appendix
{\noindent For a general bimolecular chemical reaction,
\begin{align}\label{genrxn}
  \ce{aA + bB <=>[k_f][k_r] cC + dD},
\end{align}
the concentration of the reactant A will evolve (in the absence of dynamics), according to,
\begin{equation}\label{Arate}
  \frac{\partial[\text{A}]}{\partial t}  =
  \text{k}_\text{r}[\text{C}]^\text{c}[\text{D}]^\text{d}
  -\text{k}_\text{f}[\text{A}]^\text{a}[\text{B}]^\text{b}
  +\sum_{i=0}^{N_\text{rxn}}(\mathcal{P}_i-\mathcal{L}_i),
\end{equation}
where $\mathcal{P}_i$ and $\mathcal{L}_i$ are the production and loss rates for other reactions in
the chemical network.
%which, in the case of a chemical reaction network generalizes to,
%\begin{equation}
%  \frac{\partial[\text{A}]}{\partial t}  = \sum_{i=1}^{N_\text{rxn}}
%    \text{k}_{\text{r},i}[\text{C}_i]^{\text{c}_i}[\text{D}_i]^{\text{d}_i}
%  -\text{k}_{\text{f},i}[\text{A}]^{\text{a}_i}[\text{B}_i]^{\text{b}_i}
%\end{equation}
%Chemical equilibrium is achieved when the rate of the forward reaction is equal to the rate of the
%reverse reaction,
%\begin{equation}
%  \text{k}_\text{f}[\text{A}]_\text{eq}^\text{a}[\text{B}]_\text{eq}^\text{b} =
%  \text{k}_\text{r}[\text{C}]_\text{eq}^\text{c}[\text{D}]_\text{eq}^\text{d},
%\end{equation}
%which can be used to rewrite Equation \ref{Arate} as,
%\begin{equation}\label{Arate_sub}
%  \frac{\partial[\text{A}]}{\partial t}  =
%  \text{k}_\text{f}\left(\frac{[\text{C}]^\text{c}[\text{D}]^\text{d}}
%       {[\text{C}]_\text{eq}^\text{c}[\text{D}]_\text{eq}^\text{d}}
%       [\text{A}]_\text{eq}^\text{a}[\text{B}]_\text{eq}^\text{b}
%       -[\text{A}]^\text{a}[\text{B}]^\text{b}\right)
%\end{equation}
If the forward reaction given by Equation \ref{genrxn} is the rate-limiting step for the destruction or
production of a species X, then
the timescale for a system to relax to equilibrium from some initial concentration of X,
[X]$_0$, can be estimated as,
\begin{equation}\label{Arelax}
  \frac{\Delta[\text{X}]}{\tau_\text{chem}} \approx
  -\text{k}_\text{f}[\text{A}]^\text{a}[\text{B}]^\text{b}
  \rightarrow \tau_\text{chem} \approx \frac{[\text{X}]_\text{eq}-[\text{X}]_0}{-\text{k}_\text{f}[\text{A}]^\text{a}[\text{B}]^\text{b}}.
\end{equation}
where the assumption is made that the reverse reaction is very slow compared to the forward reaction and
that the forward reaction is part of the only significant mechanism for X production/destruction.
For a species to quench, an atmospheric parcel dynamically perturbed from equilibrium must be replenished
by the dynamics on a timescale, $\tau_\text{dyn}$, shorter than this relaxation timescale,
$\tau_\text{chem}$. Following PB77, the destruction or production rate of X above the quench point
(where $\tau_\text{chem}=\tau_\text{dyn}$)
can be written as,
\begin{equation}
\frac{\Delta X}{\tau_\text{chem}} = \frac{\Delta X(z=0)}{\tau_\text{chem}(z=0)}\exp\left(-\frac{z}{H_\text{chem}}\right),
\end{equation}
where $z=0$ is defined as the quench point, and the chemical scale height is defined as,
\begin{equation}
  H_\text{chem} =
  -\left[\frac{\partial}{\partial z}\ln\left(\Delta X/\tau_\text{chem}\right)\right]^{-1}
  =
  -\left[\frac{\partial}{\partial z}\ln\left(\text{k}_\text{f}[\text{A}]^\text{a}[\text{B}]^\text{b}\right)\right]^{-1}.
\end{equation}
This expression can be simplified in terms of a series of scale heights,
\begin{equation}
  H_\text{chem} = \biggl[H_{\text{k}_\text{f}}^{-1} + (\text{a+b})H_n^{-1}-\left(\text{a}\frac{\partial}{\partial z}\ln f_\text{A}\right)^{-1}-\left(\text{b}\frac{\partial}{\partial z}\ln f_\text{B}\right)^{-1}\biggr]^{-1}
\end{equation}
where $H_{\text{k}_\text{f}}$ and $H_n$ are the scale heights of the forward rate constant and number density, and $f_\text{A}$ and $f_\text{B}$ are the mole fractions of A and B.}

{
In our work, we assume that the species X and A are the same species, which we have called c, as
described in the reaction in Equation \ref{crxn}. Further, we have assumed that the second reactant is the
background gas, such that the mole fraction $f_B$ is essentially constant with height. This leads to the
expression for the full chemical scale height given by Equation \ref{full_chemical_scaleheight}.
}
\bibliography{diseq_bib}
\listofchanges



\end{document}